\documentclass{phb-proc4-auth}

\usepackage{graphicx}
\usepackage{amssymb}

\begin{document}
\begin{frontmatter}


\journal{SCES '04}


\title{$F$-electron spectral function near a quantum critical point}

%
%
%
%
%
%

\author[US]{J.~K.~Freericks\corauthref{1}}
\author[US]{V.~M.~Turkowski}
\author[HR]{V.~Zlati\'c}

%
 
\address[US]{Department of Physics, Georgetown University, Washington, DC
20057, USA}
\address[HR]{Institute of Physics, Bijenicka c.~46, P.~O.~B.~304, 10000 
Zagreb, Croatia}

%
%
%
%


%
%
%
%

\corauth[1]{Corresponding Author: Department of Physics, Georgetown University,
Washington, DC 20057, USA. Phone: (202) 687-6159, Fax: (202) 687-2087,
Email: freericks@physics.georgetown.edu}


\begin{abstract}
We calculate the $f$-electron spectral function using a Keldysh formalism
for the Falicov-Kimball model in infinite dimensions.  We study the region
close to the quantum critical point on both the hypercubic and
Bethe lattices.
\end{abstract}

%
%



\end{frontmatter}

%
%
%
%
%
The Falicov-Kimball model~\cite{falicov_kimball} involves  
conduction electrons, which are free to move through the 
lattice, and $f$-electrons which are immobile.  The two electrons
interact with each other by a Coulomb interaction (of strength $U$) when
they are located at the same lattice site.  The Hamiltonian is (at half filling)
\begin{eqnarray}
\mathcal{H}&=&-\frac{t^*}{\sqrt{Z}}\sum_{\langle ij \rangle} (c^\dagger_ic_j
+c^\dagger_jc_i)
+U\sum_i c^\dagger_ic_if^\dagger_if_i\nonumber \\
&-&\frac{U}{2}\sum_i(c^\dagger_ic_i
+f^\dagger_if_i),
\label{eq: ham_def}
\end{eqnarray}
where $c^\dagger_i$ ($c_i$) creates (destroys) a conduction electron at
site $i$, $f^\dagger_i$ ($f_i$) creates (destroys) a localized electron at
site $i$, and $t^*$ is the hopping integral~\cite{metzner_vollhardt}.
The symbol $Z$ represents the number of nearest neighbors, and
$\langle i j\rangle$ denotes a sum over all nearest neighbor pairs.
The formalism for solving the conduction-electron Green's function
was worked out by Brandt and Mielsch~\cite{brandt_mielsch}.

\begin{figure}
\begin{center}
\includegraphics[width=1.25in]{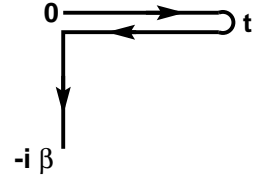}
\caption{\label{fig: contour} Keldysh contour for evaluating the $f$-electron
Green's function at time $t$.  The contour runs from $0$ to $t$,
then back from $t$ to $0$ and finally goes
along the imaginary axis down to  $-i\beta$.  
When we discretize the matrix operator over the
Keldysh contour, we evaluate the integrals via a rectangular (midpoint)
summation.  We typically use no more than 2500 time steps on
the contour.}
\end{center}
\end{figure}

Brandt and Urbanek~\cite{brandt_urbanek} describe how to
calculate the $f$-electron spectral function using the Keldysh technique  
to directly determine the greater Green's function (along the Keldysh contour)
and then Fourier transforming to real frequency (see also~\cite{fk_review} for
a review of the formalism).  The greater Green's function is defined to be
\begin{equation}
G_f^>(t)=-\textrm{Tr}\langle e^{-\beta \mathcal{H}_{imp}}S_c(\lambda)f(t)
f^\dagger(0)\rangle/\mathcal{Z}_{imp},
\label{eq: green_f_def}
\end{equation}
with $f(t)=\exp(it\mathcal{H}_{imp})f\exp(-it\mathcal{H}_{imp})$ and
the evolution operator is given by
\begin{equation}
S_c(\lambda)=\mathcal{T}_c\exp\left [ \int_cd\bar t\int_c d\bar t^\prime
c^\dagger(\bar t)\lambda_c(\bar t, \bar t^\prime )c(\bar t^\prime )\right ] .
\label{eq: s_lambda_c}
\end{equation}
The subscript $imp$ denotes the use of the impurity Hamiltonian (no hopping)
with the evolution
operator corresponding to the dynamical mean field~\cite{brandt_mielsch} 
$\lambda(\omega)$ (the dynamical mean field mimics the hopping of conduction
electrons onto and off of a given site; it is adjusted so that the impurity
conduction-electron Green's function equals the local lattice Green's function).
The time-ordering is along the Keldysh contour (see Fig.~\ref{fig: contour}),
and the contour-ordered
dynamical mean field is found from a Fourier transform of $\lambda(\omega)$
\begin{eqnarray}
\lambda_c(\bar t, \bar t^\prime )&=&-\frac{1}{\pi}\int_{-\infty}^\infty
d\omega \textrm{Im} \lambda(\omega)\exp[-i\omega(\bar t -\bar t^\prime )]\nonumber\\
&\times&[f_{FD}(\omega)-\theta_c(\bar t - \bar t^\prime )],
\label{eq: lambda_c}
\end{eqnarray}
where $f_{FD}(\omega)=1/[1+\exp(\beta\omega)]$ is the Fermi-Dirac distribution
and $\theta_c(\bar t - \bar t^\prime )=0$ if $\bar t^\prime$ is in front of
$\bar t$ on the contour $c$ and 1 if it is behind. The Green's function
can be solved by directly evaluating the Feynman path integral over the contour,
which reduces to calculating the determinant of a continuous matrix operator.
This operator is discretized, and a conventional determinant is evaluated
when we compute results numerically. Once $G^>(t)$ is found, then the DOS 
(at half filling) is determined by 
\begin{equation}
A_f(\omega)=-\frac{2}{\pi}\int_0^\infty dt \textrm{Re}\{G_f^>(t)\}
\cos(\omega t).
\label{eq: f_spect_final}
\end{equation}

We work with two different infinite coordination 
number lattices 
here: (i) the Bethe lattice, which has a noninteracting DOS that is
a semicircle $\rho_B(\omega)=\sqrt{4t^{*2}-\omega^2}/2\pi t^{*2}$
and (ii) the hypercubic lattice, which has a noninteracting 
DOS that is a Gaussian $\rho_H(\omega)=\exp(-\omega^2/t^{*2})/\sqrt{\pi}t^*$.
All energies are measured in units of $t^*$.

\begin{figure}[htbf]
\begin{center}
\includegraphics[width=2.55in]{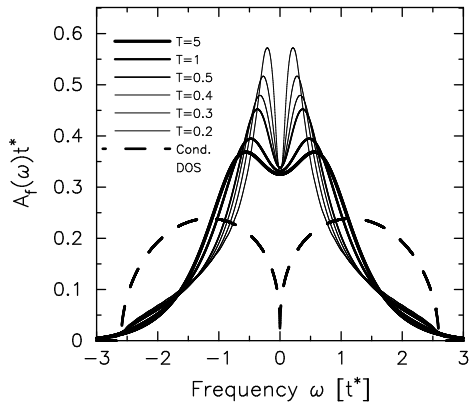}
\caption{
$F$-electron DOS for the Bethe lattice.  The different thickness lines 
correspond to different temperatures.
The conduction electron DOS is the dashed line. 
\label{fig: f_dos_bethe}}
\end{center}
\end{figure}

\begin{figure}[htbf]
\begin{center}
\includegraphics[width=2.55in]{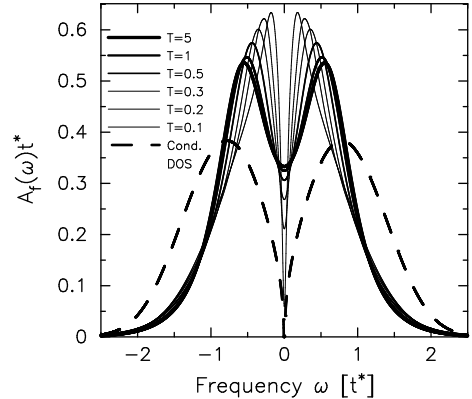}
\caption{
Similar plot of the $f$-electron DOS for the hypercubic lattice.
\label{fig: f_dos_hc}}
\end{center}
\end{figure}

In Fig.~\ref{fig: f_dos_bethe}, we plot the DOS at the critical value of $U$
for the Mott transition on the Bethe lattice ($U=2$), and in 
Fig.~\ref{fig: f_dos_hc} just on the insulating
side of the transition on the hypercubic lattice ($U=1.5$).  Also included is
a plot of the conduction-electron DOS, which is independent of 
temperature~\cite{vandongen}.  Note how the $f$-electron spectral function
has an interesting temperature evolution, where it develops a pseudogap
as $T$ is reduced.  Calculations become more difficult at lower
temperatures~\cite{numerical} (because of discretization errors
along the Keldysh contour), and rapidly exhaust our computational resources.
As a check on the accuracy, we use the DOS to calculate the Matsubara frequency
Green's functions and compare them to results calculated directly on the 
imaginary axis, and we compare the first three moments 
to the exact results for those moments.
In general, we need to extrapolate the discretization size to zero to get 
accurate results, and most reported DOS here are accurate to at least 1\% in
all of the moments and 0.1\% for all Matsubara frequency Green's functions.
A long paper evaluating different values of $U$ is in 
preparation~\cite{numerical}.

\textit{Acknowledgments}: We 
acknowledge support of the Office of Naval Research 
(N00014-99-1-0328) and from the National Science Foundation 
(DMR-0210717).  Supercomputer resources were provided by
the Arctic Region Supercomputer Center and the Engineering Research
and Development Center.
%
%
%
%

%
%
%
%


\end{document}